\documentclass[aip,amsmath,amssymb,reprint,]{revtex4-1}

\usepackage{graphicx}% Include figure files
\usepackage{dcolumn}% Align table columns on decimal point
\usepackage{bm}% bold math
\usepackage[utf8]{inputenc}
\usepackage[T1]{fontenc}
\usepackage{mathptmx}
\usepackage{etoolbox}
\usepackage{xcolor}
\usepackage{hyperref}

\usepackage{verbatim}

\makeatletter
\def\@email#1#2{%
 \endgroup
 \patchcmd{\titleblock@produce}
  {\frontmatter@RRAPformat}
  {\frontmatter@RRAPformat{\produce@RRAP{*#1\href{mailto:#2}{#2}}}\frontmatter@RRAPformat}
  {}{}
}%
\makeatother

\begin{document}

% Don't count these!
%TC:ignore
%\quickwordcount{aipsamp}
%\quickcharcount{aipsamp}
%\detailtexcount{aipsamp} %<- this one
%TC:endignore

\preprint{AIP/123-QED}

%TC:ignore
\title[Low Energy Backgrounds and Excess Noise in a Two-Channel Low-Threshold Calorimeter]{Low Energy Backgrounds and Excess Noise in a Two-Channel Low-Threshold Calorimeter}
% Force line breaks with \\

\author{R. Anthony-Petersen} \affiliation{University of California Berkeley, Department of Physics, Berkeley, CA 94720, USA}
%\author{A. Biekert} \affiliation{University of California Berkeley, Department of Physics, Berkeley, CA 94720, USA}
%\author{H. Birch} \affiliation{University of Michigan, Randall Laboratory of Physics, Ann Arbor, MI 48109-1040, USA}
%\author{T.K. Bui} \affiliation{International Center for Quantum-field Measurement Systems for Studies of the Universe and Particles (QUP,WPI), High Energy Accelerator Research Organization (KEK), Oho 1-1, Tsukuba, Ibaraki 305-0801, Japan}
\author{C.L. Chang} \affiliation{Argonne National Laboratory, 9700 S Cass Ave, Lemont, IL 60439, USA} \affiliation{Kavli Institute for Cosmological Physics, The University of Chicago, Chicago, IL 60637, USA} \affiliation{Department of Astronomy and Astrophysics, The University of Chicago, Chicago, IL 60637, USA}
\author{Y.-Y. Chang} \affiliation{University of California Berkeley, Department of Physics, Berkeley, CA 94720, USA}
\author{L. Chaplinsky} \affiliation{University of Massachusetts, Amherst Center for Fundamental Interactions and Department of Physics, Amherst, MA 01003-9337, USA}
%\author{G. Cline} \affiliation{Lawrence Berkeley National Laboratory, 1 Cyclotron Rd., Berkeley, CA 94720, USA}
%\author{A. Dushkin} \affiliation{University of Michigan, Randall Laboratory of Physics, Ann Arbor, MI 48109-1040, USA}
\author{C.W. Fink} \affiliation{University of California Berkeley, Department of Physics, Berkeley, CA 94720, USA} \affiliation{Now at Los Alamos National Laboratory, Los Alamos, NM 87545, USA}
%\author{S. Fiorucci} \affiliation{Lawrence Berkeley National Laboratory, 1 Cyclotron Rd., Berkeley, CA 94720, USA}
\author{M. Garcia-Sciveres} \affiliation{Lawrence Berkeley National Laboratory, 1 Cyclotron Rd., Berkeley, CA 94720, USA} \affiliation{International Center for Quantum-field Measurement Systems for Studies of the Universe and Particles (QUP,WPI), High Energy Accelerator Research Organization (KEK), Oho 1-1, Tsukuba, Ibaraki 305-0801, Japan}
%\author{G. Gilchriese} \affiliation{Lawrence Berkeley National Laboratory, 1 Cyclotron Rd., Berkeley, CA 94720, USA}
\author{W. Guo} \affiliation{Department of Mechanical Engineering, FAMU-FSU College of Engineering, Florida State University, Tallahassee, FL 32310, USA} \affiliation{National High Magnetic Field Laboratory, Tallahassee, FL 32310, USA}
\author{S.A. Hertel} \affiliation{University of Massachusetts, Amherst Center for Fundamental Interactions and Department of Physics, Amherst, MA 01003-9337, USA}
%\author{A. Jastram} \affiliation{Texas A\&M University, Department of Physics and Astronomy, College Station, TX 77843-4242, USA}
\author{X. Li} \affiliation{Lawrence Berkeley National Laboratory, 1 Cyclotron Rd., Berkeley, CA 94720, USA}
\author{J. Lin} \affiliation{University of California Berkeley, Department of Physics, Berkeley, CA 94720, USA} \affiliation{Lawrence Berkeley National Laboratory, 1 Cyclotron Rd., Berkeley, CA 94720, USA}
\author{M. Lisovenko} \affiliation{Argonne National Laboratory, 9700 S Cass Ave, Lemont, IL 60439, USA}
\author{R. Mahapatra} \affiliation{Texas A\&M University, Department of Physics and Astronomy, College Station, TX 77843-4242, USA}
\author{W. Matava} \affiliation{University of California Berkeley, Department of Physics, Berkeley, CA 94720, USA}
\author{D.N. McKinsey} \affiliation{University of California Berkeley, Department of Physics, Berkeley, CA 94720, USA} \affiliation{Lawrence Berkeley National Laboratory, 1 Cyclotron Rd., Berkeley, CA 94720, USA}
\author{D.Z. Osterman} \affiliation{University of Massachusetts, Amherst Center for Fundamental Interactions and Department of Physics, Amherst, MA 01003-9337, USA}
\author{P.K. Patel} \affiliation{University of Massachusetts, Amherst Center for Fundamental Interactions and Department of Physics, Amherst, MA 01003-9337, USA}
\author{B. Penning} \affiliation{University of Zurich, Department of Physics, 8057 Zurich, Switzerland}
%\author{H.D. Pinckney} \affiliation{University of Massachusetts, Amherst Center for Fundamental Interactions and Department of Physics, Amherst, MA 01003-9337 USA}
\author{M. Platt} \affiliation{Texas A\&M University, Department of Physics and Astronomy, College Station, TX 77843-4242, USA}
\author{M. Pyle} \affiliation{University of California Berkeley, Department of Physics, Berkeley, CA 94720, USA}
\author{Y. Qi} \affiliation{Department of Mechanical Engineering, FAMU-FSU College of Engineering, Florida State University, Tallahassee, FL 32310, USA} \affiliation{National High Magnetic Field Laboratory, Tallahassee, FL 32310, USA}
\author{M. Reed} \affiliation{University of California Berkeley, Department of Physics, Berkeley, CA 94720, USA}
\author{I. Rydstrom} \affiliation{University of California Berkeley, Department of Physics, Berkeley, CA 94720, USA}
%\author{G.R.C Rischbieter} \affiliation{University of Michigan, Randall Laboratory of Physics, Ann Arbor, MI 48109-1040, USA}
\author{R.K. Romani} \thanks{Corresponding author: \href{mailto:rkromani@berkeley.edu}{rkromani@berkeley.edu}}\affiliation{University of California Berkeley, Department of Physics, Berkeley, CA 94720, USA}
\author{B. Sadoulet}\affiliation{University of California Berkeley, Department of Physics, Berkeley, CA 94720, USA}
\author{B. Serfass} \affiliation{University of California Berkeley, Department of Physics, Berkeley, CA 94720, USA}
%\author{R.J.  Smith} \affiliation{University of California Berkeley, Department of Physics, Berkeley, CA 94720, USA}
\author{P. Sorensen} \affiliation{Lawrence Berkeley National Laboratory, 1 Cyclotron Rd., Berkeley, CA 94720, USA}
\author{B. Suerfu} \affiliation{International Center for Quantum-field Measurement Systems for Studies of the Universe and Particles (QUP,WPI), High Energy Accelerator Research Organization (KEK), Oho 1-1, Tsukuba, Ibaraki 305-0801, Japan}
%\author{A. Suzuki} \affiliation{Lawrence Berkeley National Laboratory, 1 Cyclotron Rd., Berkeley, CA 94720, USA}
\author{V. Velan} \affiliation{Lawrence Berkeley National Laboratory, 1 Cyclotron Rd., Berkeley, CA 94720, USA}
\author{G. Wang} \affiliation{Argonne National Laboratory, 9700 S Cass Ave, Lemont, IL 60439, USA}
\author{Y. Wang} \affiliation{University of California Berkeley, Department of Physics, Berkeley, CA 94720, USA}
\author{S.L. Watkins} \affiliation{University of California Berkeley, Department of Physics, Berkeley, CA 94720, USA}
\author{M.R. Williams} \affiliation{Lawrence Berkeley National Laboratory, 1 Cyclotron Rd., Berkeley, CA 94720, USA}
%\author{V.G. Yefremenko} \affiliation{Argonne National Laboratory, 9700 S Cass Ave, Lemont, IL 60439, USA}

\collaboration{TESSERACT Collaboration}%\noaffiliation

 \email{rkromani@berkeley.edu.}

\date{\today}% It is always \today, today,
             %  but any date may be explicitly specified

\begin{abstract}
We describe observations of low energy excess (LEE) events, background events observed in all light dark matter direct detection calorimeters, and noise in a Transition Edge Sensor based two-channel silicon athermal phonon detector with 375 meV baseline energy resolution. We measure two distinct LEE populations: ``shared'' multichannel events with a pulse shape consistent with substrate athermal phonon events, and sub-eV events that couple nearly exclusively to a single channel with a significantly faster pulse shape. These ``singles'' are consistent with events occurring within the aluminum athermal phonon collection fins. Similarly, our measured detector noise is higher than the theoretical expectation. Measured noise can be split into an uncorrelated component, consistent with shot noise from small energy depositions within the athermal phonon sensor itself, and a correlated component, consistent with shot noise from energy depositions within the silicon substrate's phonon system. 
\end{abstract}

\maketitle

%\section{Introduction}
%TC:endignore

The search for dark matter (DM) has expanded to lower mass candidates, including sub-GeV ``light mass'' DM \cite{kuflikElasticallyDecouplingDark2016,kuflikPhenomenologyELDERDark2017a, hochbergMechanismThermalRelic2014,hochbergModelThermalRelic2015, hallFreezeinProductionFIMP2010}. Direct detection of light mass DM scattering off nuclei, electrons, or crystal lattices requires extremely low energy thresholds, given the low kinetic energy carried by the DM particles. Cryogenic calorimeters are well suited to attaining such low thresholds, and have recently set limits on sub-GeV DM-nucleon interaction cross sections \cite{cresstcollaborationFirstResultsCRESSTIII2019, CPDLimits, CRESST2023Limits}. These calorimeters typically read out athermal phonons from a crystalline substrate using Transition Edge Sensors (TES) \cite{irwinTransitionEdgeSensors2005} connected to aluminum athermal phonon collection fins (forming structures known as Quasiparticle-trap-assisted Electrothermal-feedback TESs: QETs \cite{irwinQuasiparticleTrapAssisted1995}). 

Calorimetric DM direct detection experiments and other low threshold calorimeters have observed an excess of events below several hundred eV, with a rate that rises dramatically at low energies \cite{adariEXCESSWorkshopDescriptions2022, CPDLimits, cresstcollaborationFirstResultsCRESSTIII2019}. The rate of these low energy excess (LEE) events decreases with time, and can be regenerated by warming up the detector \cite{angloherLatestObservationsLow2022}. Additionally, the LEE rate varies only weakly with detector material or mass \cite{angloherLatestObservationsLow2022}, and appears similarly in detectors run above and below ground\cite{PyleEXCESS2022}.

The decrease in LEE rate with time suggests a relaxation mediated process. Mechanical stress relaxation in the detector holding has been shown to create LEE-like events \cite{astromFractureProcessesObserved2006, AnthonyPetersen2024}; however, a LEE population remains even in detectors held in low stress configurations \cite{AnthonyPetersen2024} (when not discriminating between ``singles'' and ``shared'' LEE, as we do here), implying additional relaxation processes are necessary to explain observations.

Stress created by the thermal contraction of sensor films relative to thick detector substrates has been proposed as another LEE source \cite{AnthonyPetersen2024}. This stress would be present in all calorimeters observing the LEE, largely independent of the detector material or size.

If relaxation within films is responsible for the LEE, we expect some partitioning of energy between local deposition in the film's electronic system and phonon energy that leaks into the substrate. For example, athermal phonon bursts within the aluminum QET should break Cooper pairs and locally deposit energy \cite{KozorezovQPs, MartinisSavingQubits}. A detector instrumented with multiple individually-read-out athermal phonon sensors would measure this localized energy deposition and therefore be able to discriminate LEE events (with significant local energy absorption) from DM interactions in the bulk substrate (that approximately uniformly excite all phonon sensors). This detector architecture has previously been shown by CRESST\cite{CRESSTTwoChannel} and contemporaneously by our group \cite{TwoChannelTalkLTD} to have great potential. Around this time, a version of the film relaxation model briefly proposed in Ref. \cite{AnthonyPetersen2024} was more fully developed and published \cite{AlRelaxation}. This model attempted to explain LEE phonon bursts through the relaxation of thermally stressed aluminum films on the device surface. During the early experimental work\cite{TwoChannelTalkLTD}, limited control of systematics and understanding of pulse shapes for locally absorbed events constrained our ability to draw strong conclusions about sources of LEE. These limitations have now been remedied.

%TC:ignore
%\section{Detector Design and Data Collection}
%TC:endignore

\begin{figure}
\includegraphics[width=1\columnwidth]{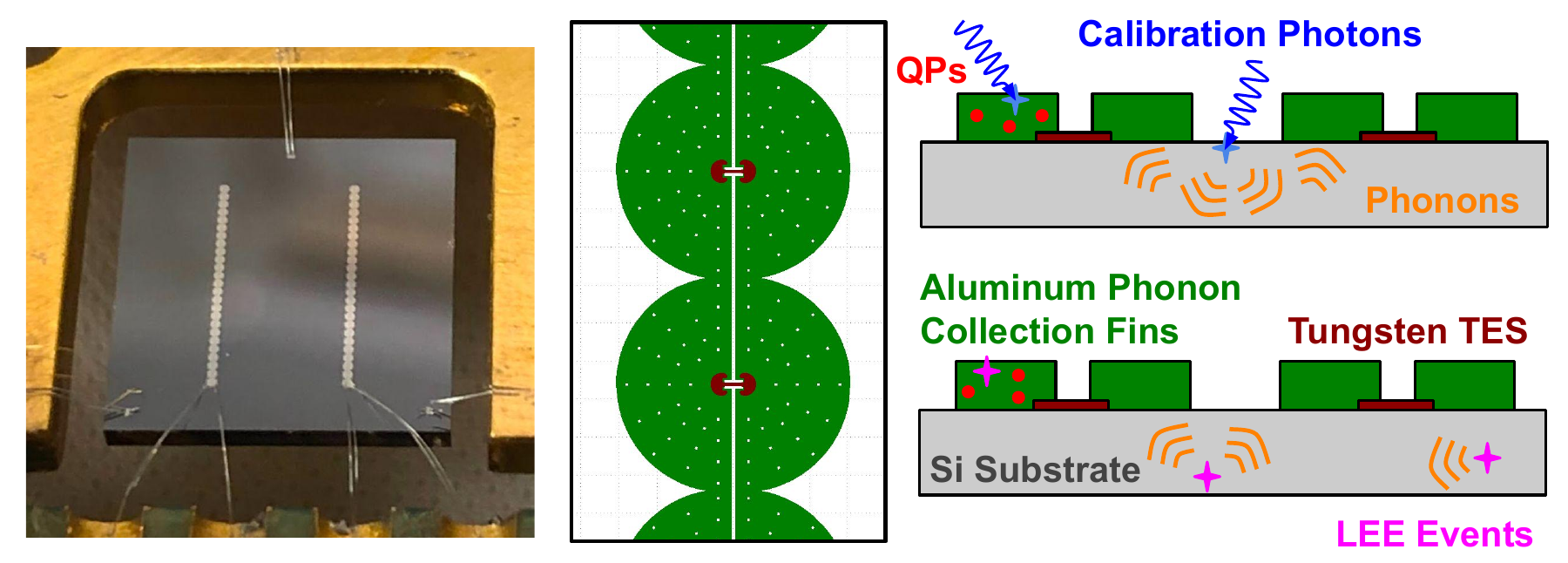}
\caption{\label{fig:diagram} (Left) Photograph of our 1 cm $\times$ 1 cm $\times$ 1 mm thick silicon athermal phonon detector. Two columns of QETs compose our phonon sensors, which are connected to a readout PCB (below) with wire bonds. The device is suspended by wire bonds as in Ref. \cite{AnthonyPetersen2024} to suppress detector holding backgrounds. A gold wire bond (left) serves to cool the device. (Center) detail of the QET\cite{irwinQuasiparticleTrapAssisted1995} design, showing aluminum phonon collection fins (green), the tungsten/aluminum overlap region (purple half circle) and tungsten TESs (purple rectangles connecting overlap regions). For scale, the QET fins are 140 $\mu$m in radius. As the aluminum is deposited over the half moon shaped W/Al overlap region, only the thin rectangular W TES is visible to calibration photons from above. 25 quasi-circular QETs are wired in parallel to form a channel. (Right Top) Sketch of the interaction of calibration photons with the detector, creating phonon (orange) or quasiparticle (QP, red) bursts in the detector substrate or aluminum QET fins respectively. (Right Bottom) Sketch of LEE events, causing both phonon bursts in the detector substrate, and quasiparticle bursts in the aluminuim QET fins.} 
\end{figure}

To test this LEE discrimination concept, a 1 cm-square, 1 mm-thick silicon substrate was instrumented with two channels of 25 tungsten TESs ($\sim$ 48 mK Tc, nominally 40 nm thick) connected to aluminum athermal phonon collection fins (QETs \cite{irwinQuasiparticleTrapAssisted1995}, 600 nm thick), covering 1.38$\%$ of the device's surface (see Fig. \ref{fig:diagram}). QETs were electrically connected by partially overlapping their aluminum fins, such that each channel of 25 TESs was read out as one unit. Unfortunately, during manufacturing, a fraction of the TESs were partially etched away, leading to some performance degradation (higher normal resistance and worse phonon collection efficiency). See supplementary material section C and Fig. S4 for further discussion. While performance in both channels was still acceptable, we focus on the left channel, which was less negatively impacted. As in Ref. \cite{AnthonyPetersen2024}, this detector was suspended from wire bonds to minimize LEE-type backgrounds associated with detector holding. It was housed inside multiple layers of electromagnetic interference (EMI) and infrared (IR) shielding at the base stage of a dilution refrigerator, and was read out using DC SQUID array amplifiers.

\begin{figure}
\includegraphics[width=1\columnwidth]{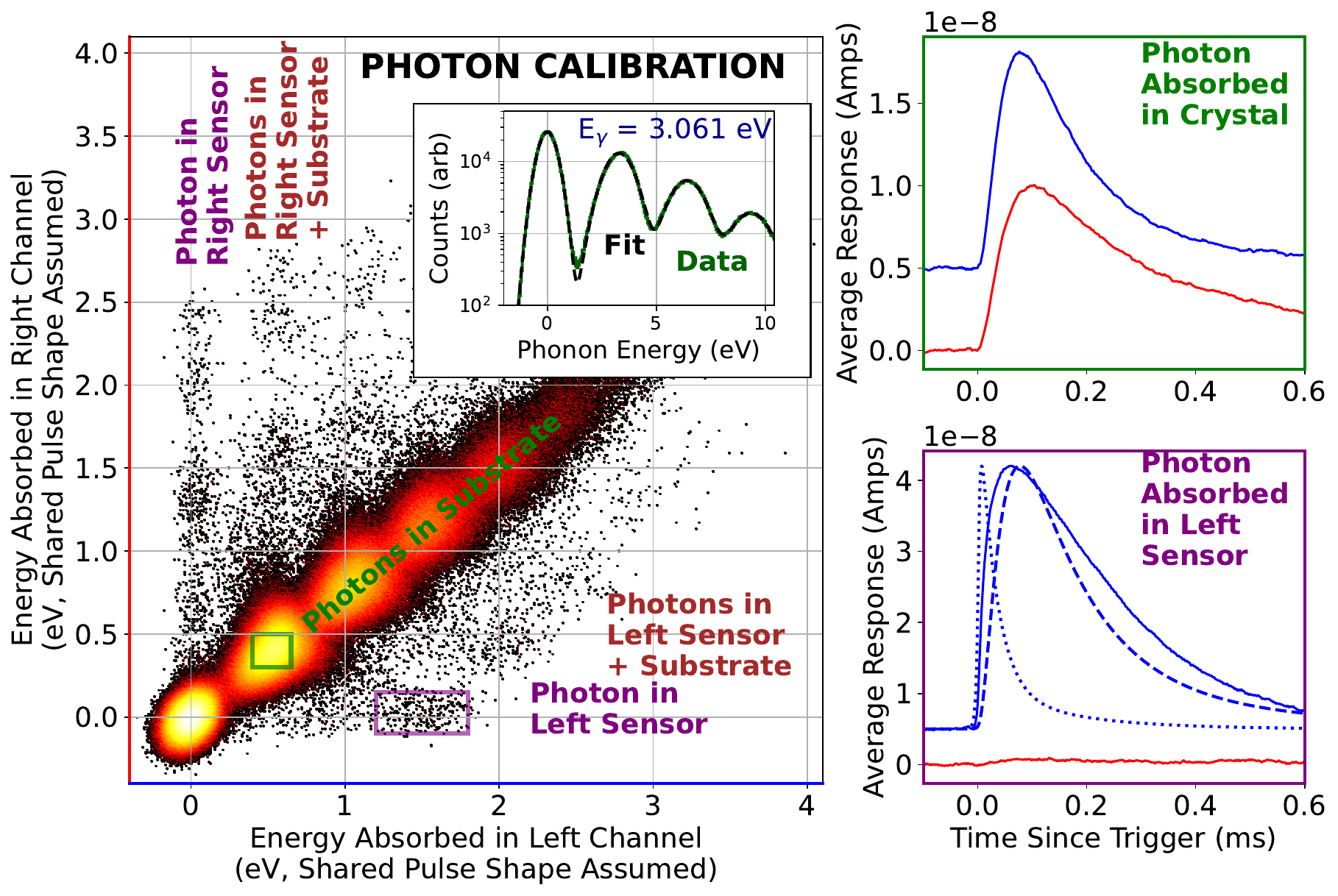}
\caption{\label{fig:calibration} (Left) Two dimensional histogram of energy absorbed in the left (blue) and right (red) channels on the detector during photon calibration, assuming a phonon-like pulse shape. (Inset) Histogram of combined phonon energies (see supplementary material section C for further discussion of this combination technique), with multi-Gaussian fit (dashed). (Right) Average response for substrate (top; green box in left panel) and direct hit (bottom; purple box in left panel) events. Traces are filtered above 50 kHz and offset vertically for clarity. Solid red and blue correspond to right and left channel responses. Dashed line shows the phonon template, while dotted line shows the modeled TES response to Dirac delta impulses. Note that energy reconstruction in this plot assumes a pulse shape for substrate/shared/phonon events, and direct hit events will not be reconstructed at the correct energy. See supplementary material section C for further discussion of our energy reconstruction.}
\end{figure}

The detector was calibrated with optical photons to characterize its response to events of a known energy (see Figure \ref{fig:calibration}). Pulses of photons from a 405 nm (3.061 eV) room temperature laser were transmitted to the device using a single mode optical fiber terminated with a diffuser, which dispersed  photons across the entire instrumented side of the detector. The photon pulses were fast ($\sim$ 1 $\mu$s) compared to the electrical ($\sim$ 10 $\mu$s) and electrothermal ($\sim$ 100 $\mu$s) response times of the TESs. On average $\sim 0.76$ photons hit the detector per pulse. We recorded calibration data for 3.5 hours, firing laser pulses at 100 Hz, ultimately recording 1.26 million calibration events. Immediately following the calibration, three hours of background data were acquired, interleaved with periods where the TESs were characterized using IV and dIdV measurements \cite{irwinTransitionEdgeSensors2005, watkinsThesis}.

Both datasets were recorded continuously and triggered offline. For the calibration dataset, we triggered on a recorded logic signal in coincidence with the laser pulses. In the background dataset, events were triggered on the sum of the two detector channels using an optimum filter energy estimator \cite{sunilThesis}. See supplementary material section A for further discussion of triggering.

\begin{figure}
\includegraphics[width=1\columnwidth]{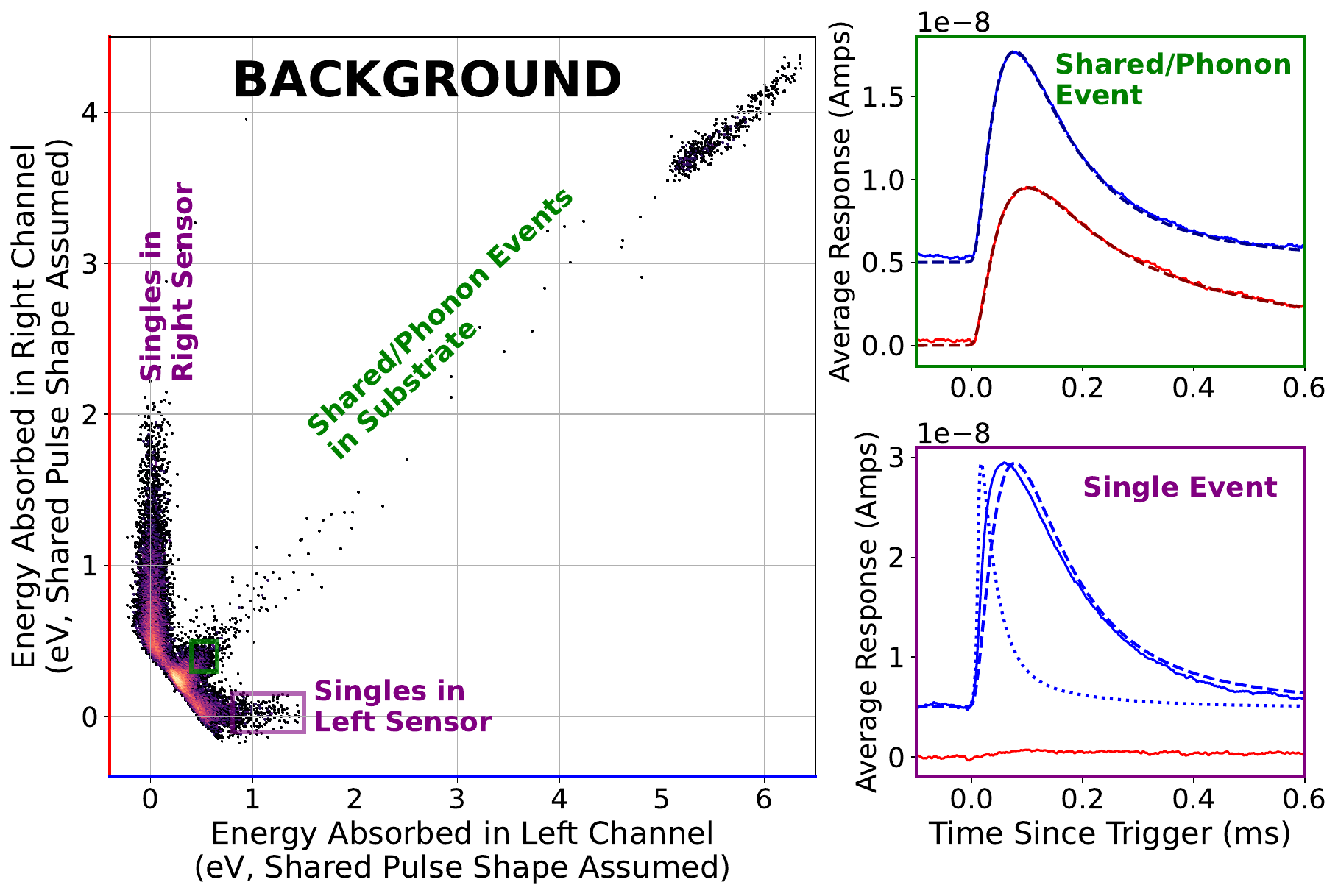}
\caption{\label{fig:backgrounds} (Left) Energies of background events in the left and right channels of the detector, assuming a phonon pulse shape and triggering on the sum of both channels. The group of events above 5 eV in the Left channel are saturated events, consistent with cosmic rays and radiogenic backgrounds (see supplementary material section D additional discussion).
(Top right) shows the average responses for events in the green box (shared events) which match phonon templates derived from the photon calibration (dashed). (Bottom right) The averaged pulse shape for ``single'' events (purple box). Dashed line: phonon template, dotted line: modeled TES response to Dirac delta impulses. Traces are filtered above 50 kHz and offset vertically for clarity. Solid red and blue correspond to right and left channel responses.  Note that energy reconstruction in this plot assumes a pulse shape for shared events, and singles will not be reconstructed at the correct energy. See supplementary material section C for further discussion of our energy reconstruction.}
\end{figure}

Standard quality cuts were applied to the triggered data to ensure that the detector was operating stably, and to reject periods of high environmental noise or abnormal device performance. These cuts were designed for high passage of randomly triggered events (79.5 \%) and for similar passage of high energy events to minimize selection biases. See supplementary material section B for additional discussion. The height of each event was fit with an optimal filter assuming a calibrated phonon pulse shape (see below), and was converted into an energy by applying a factor derived from the channels responsivity $\partial P/\partial I(f)$ \cite{TESVeto} modeled from the measured complex impedance $\partial V/ \partial I(f)$ \cite{ThreePoledPdI, finkTES}. Using this method, we estimate the amount of energy each event deposits in a given TES channel for an assumed pulse shape (rather than into the detector phonon system as a whole). We use this as the primary quantity we plot in Figs. \ref{fig:calibration}, \ref{fig:backgrounds} and \ref{fig:backgroundspectra} as a subclass of events we observe clearly do not couple through the detector phonon system (``singles,'' see below). See supplementary material section C for further discussion of energy reconstruction.

%TC:ignore
%\section{Calibration and Background}
%TC:endignore

Due to the strong similarities in the classes of observed events, we discuss the calibration and background datasets together.

In the calibration and background datasets (see Figs. \ref{fig:calibration} and \ref{fig:backgrounds}), ``shared'' events couple roughly equally to both channels (diagonal band, left panels). These events feature a relatively slowly rising pulse (see top right panels) which can be well-modeled as the sum of two exponential phonon pulses convolved with the detector responsivity ($\partial P/\partial I(f)$). Calibration and background events are identically shaped.

We associate these events with bursts of athermal phonons from the substrate which couple roughly equally to both channels. (Phonons couple somewhat more strongly to the left channel, i.e. have a higher ``phonon collection efficiency.'' See supplementary material section C for further discussion.) In the background dataset at low energies, we associate these ``shared events'' with non-sensor film LEE relaxation sources due to the lack of significant localized energy absorption within the channel. At high energies (group of events above 5 eV in the left channel in Fig. \ref{fig:backgrounds}), the saturated event rate is roughly consistent with the expected rate of high energy events from environmental radioactivity and cosmic rays. See supplementary material section D for additional discussion.

In the calibration, these events are caused by photons absorbed in the substrate, creating quantized (0, 1, 2... photons absorbed) bursts of athermal phonons. We combine the response in both channels using inverse variance weighting\cite{TESVeto}, constructing a phonon energy estimator, and plot a combined calibration histogram (see Fig. \ref{fig:calibration}, inset, and material section C for more details on this energy reconstruction). From this histogram, we measure a world leading baseline phonon energy resolution of $\sigma_P = 375.5 \pm 0.4$ meV (stat.), favorably comparing to previous detectors\cite{CRESSTSoS, HVeVR4}, both with $\sim 1$ eV resolution. 

Another class of events exhibits a nearly maximally asymmetric channel response (vertical and horizontal bands in Figs. \ref{fig:calibration} and \ref{fig:backgrounds}), which we call ``single'' events due to their strong coupling to a single channel. In the calibration dataset, we attribute this response to events in which one photon hits an aluminum phonon collection fin, i.e. a ``direct hit'' event. Additional photons may be absorbed in the detector substrate, creating a superposition of ``direct hit'' and ``substrate'' events and forming the structure of black bands in Fig. \ref{fig:calibration}. Of the roughly 4.1$\times 10^5$ single photon events seen in the substrate, we expect $\sim$ 1 $\%$ will hit a fin in a given channel, with $\sim$10$\%$ of these photons being absorbed\cite{AlReflectivity}. 239 events fall into the purple box in Fig. \ref{fig:calibration} (which contains roughly half of the left channel singles), in general agreement with the expected number of events. We attribute the spread in reconstructed energy to instrumental effects (e.g. position dependence within the aluminum, saturation from partial TESs etching), as well as the difference in the apparent distribution of singles energies (see supplementary material section C for further discussion). Given only 0.1$\%$ of the QET area is exposed tungsten, we do not expect to observe a significant number of tungsten direct hits.

Pulse shapes for single events are shown in the bottom right panels of Figs. \ref{fig:calibration} and \ref{fig:backgrounds}. The fast rise compared to substrate events (though somewhat slower than the modeled TES response to Dirac-delta energy impulses) indicates that the substrate phonon collection dynamics are bypassed. We attribute the slow fall of these events to saturation effects. Quasiparticles created by localized photon absorption would be expected to propagate to only a few of the 25 QETs in a channel, saturating these few TESs at energies significantly below the $\sim$ 5-6 eV of absorbed energy required to saturate all 25 QETs in a channel (see the group of saturated events at high energies in Fig. \ref{fig:backgrounds}).

\begin{figure}
\includegraphics[width=1\columnwidth]{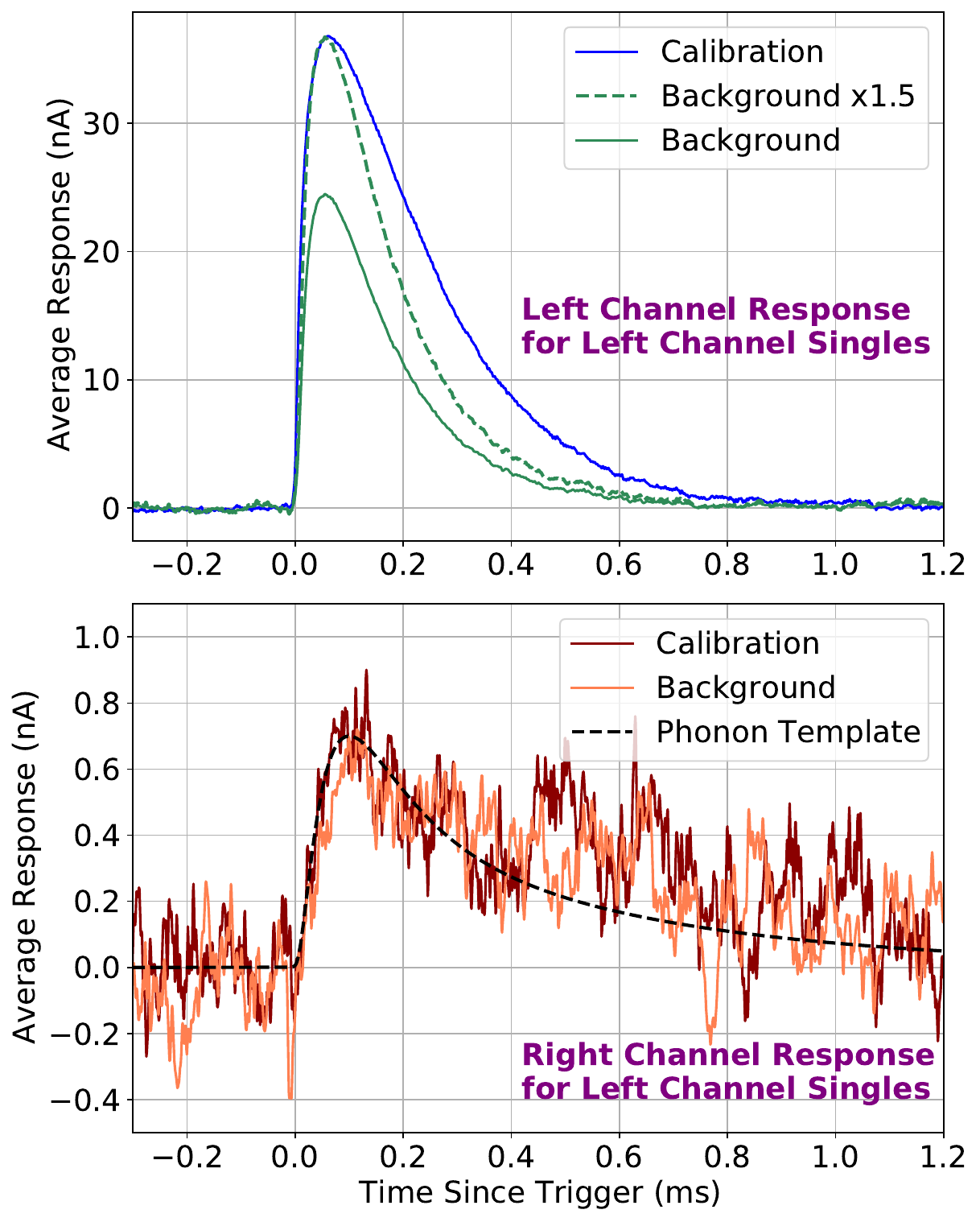}
\caption{\label{fig:pulse_shapes} Comparison between background and calibration left channel singles (purple boxed events in Figs. \ref{fig:calibration} and \ref{fig:backgrounds}). Top shows the average main (left) channel response, while bottom shows the average right channel response when a single occurs in the left channel. The top dashed line is the rescaled background singles pulse, while the bottom dashed line is the calibrated response to athermal phonon pulses. For clarity, pulses are baseline subtracted and low pass filtered at 50 kHz.}
\end{figure}

Comparing the pulse shapes of left singles in the background and calibration datasets (Fig. \ref{fig:pulse_shapes}), we see broad qualitative consistency. After rescaling, the rising edge of the background and calibration singles pulses match, while the fall for (lower energy) background singles is faster than for calibration singles. This is expected from  the saturation hypothesis: higher energy (calibration) singles would be expected to saturate more and therefore fall more slowly.

In the left singles events in Fig. \ref{fig:pulse_shapes}, we see a small average response in the opposite (right) channel, carrying a few percent of the main channel's energy, in addition to the large average response in the primary (left) channel. The average opposite (right) channel responses for background and calibration events are indistinguishable, and consistent with the calibrated athermal phonon response. We attribute this signal to athermal phonons that leak out of the aluminum fin during downconversion following a singles event. Notably, \textit{without} rescaling the average background and calibration pulse shapes seen in the opposite (right) channel, these responses are roughly equal in height, suggesting that lower energy background events are more efficient in producing athermal phonons than calibration events, and hinting that background singles may originate deeper within the aluminum fin.

In sum, we associate both calibration and background singles with events originating within an aluminum fin, locally saturating TESs close to the event and leaking out athermal phonons during the downconversion process. While calibration events are caused by photon absorption, background singles are caused by some unknown process that we broadly associate with LEE.

Bursts of high frequency (MHz-GHz) EMI \cite{TESVeto} cannot be the primary source of these background singles, as individual EMI photons would not be sufficiently energetic to leak above the aluminum gap ($2 \Delta_{\mathrm{Al}} > $ ($\sim$ 90 GHz) $\times h$) phonons into the substrate during the downconversion process. We also reject relaxation or other events occurring in the tungsten TESs as the source of background singles, as tungsten relaxation would be expected to produce a far larger pulse of athermal phonons in the substrate due to thinness of the tungsten film ($\sim 40$ nm)\cite{TESVeto}.

Photons incident on the QET fins (as in the calibration) would explain background singles. Above the silicon bandgap, these photons would also couple to the substrate, creating shared events. Future devices with improved resolutions should search for cutoffs in the shared spectrum at the silicon bandgap, characteristic of such photons.

Alternatively, the relaxation of thermally stressed aluminum films could create singles backgrounds. Romani \cite{AlRelaxation} described a model where relaxing dislocations in an aluminum film impact the aluminum-substrate interface, injecting bursts of athermal phonons into the substrate while depositing minimal energy in the aluminum (i.e. forming shared events). Clearly, observed singles are in the opposite limit: they couple almost exclusively to the aluminum film. Modifying the model in Ref. \cite{AlRelaxation} to include e.g. damping of dislocations in the bulk film through the emission of above-gap phonons or phonon-emitting interactions with intra-film grain boundaries might better explain singles events.

\begin{figure}
\includegraphics[width=1\columnwidth]{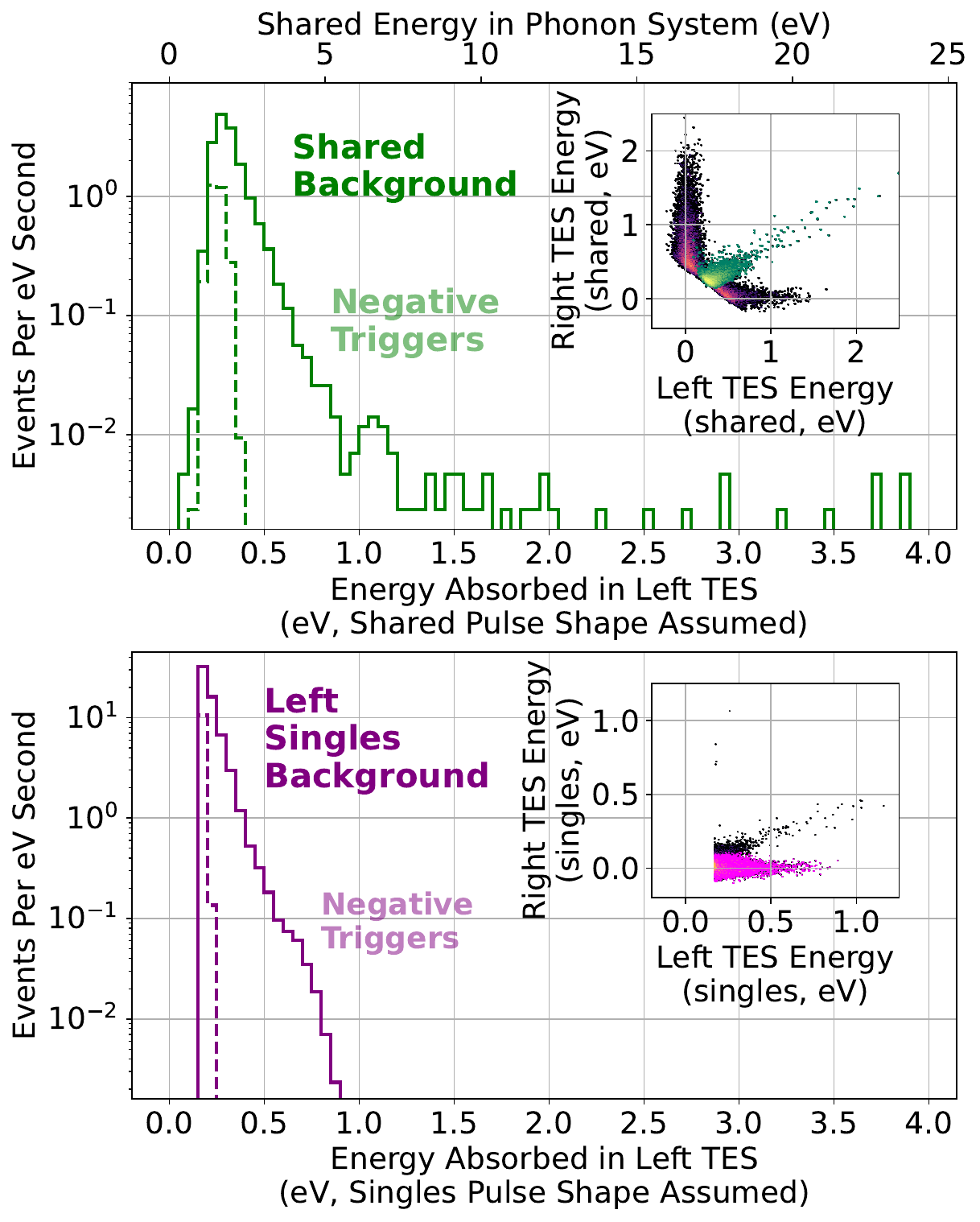}
\caption{\label{fig:backgroundspectra} Shared (green, top) and single (pink, bottom) backgrounds observed in the left channel of the detector. Dashed spectra (blue, purple) show negative amplitude events which sample noise trigger rates, demonstrating that LEE events are not predominantly noise fluctuations (see text). Inset panels show events in left vs. right energy space, with the single or shared events selected using a $\chi^2$ based approach (see text) highlighted in color. The top plots uses shared energy estimators (the top inset plot is identical to Fig. \ref{fig:backgrounds}, with shared events colored differently) and the bottom plots used singles energy estimators. See supplementary material sections A and C for more information on our triggering and energy reconstruction scheme.}
\end{figure}

The background spectra for single and shared events are plotted in Fig. \ref{fig:backgroundspectra}. Shared events were triggered on the sum of the two channels using a phonon template, while singles were triggered in the left channel using an averaged singles template. Since a given single or shared event could trigger both single and shared triggers, a $\chi^2$ statistic considering the pulse shape and amplitude in both channels was used to discriminate event types and avoid double counting. If $\chi^2_{\mathrm{single,left}} < (\chi^2_{\mathrm{shared}}, \chi^2_{\mathrm{single,right}})$, an event was classified as a left single, if $\chi^2_{\mathrm{single,right}} < (\chi^2_{\mathrm{shared}}, \chi^2_{\mathrm{single,left}})$ it was a right single, and if $\chi^2_{\mathrm{shared}} < (\chi^2_{\mathrm{single,left}}, \chi^2_{\mathrm{single,right}})$ it was determined to be a shared event. Inset plots in Fig. \ref{fig:backgroundspectra} show this $\chi^2$ based discrimination.

Above about 1 eV in the sensor ($\sim$ 6 eV in the phonon system) a slowly rising shared background dominates. At low energies, both singles and shared rates sharply and exponentially rise. While the similarities between spectra are at some level coincidental (plotting the shared channel in phonon units shows a different energy scale), we leave open the possibility that both low energy populations are caused by similar underlying processes.

To test whether the observed low energy excesses are noise artifacts, we invert the datastream and re-trigger, recording event-like noise fluctuations which were previously negative. This samples the rate of random noise events. Positive amplitude (i.e. physical) events dominate over negative amplitude events (from noise, see dashed spectra in Fig. \ref{fig:backgroundspectra}), indicating that our backgrounds are predominantly true low energy events down to the trigger threshold.

%TC:ignore
%\section{Noise Correlation}
%TC:endignore

TESs have relatively well understood noise performance, dominated by TES Johnson noise at high frequencies and Thermal Fluctuation Noise (TFN) at frequencies below the primary dynamical pole of the TES\cite{irwinTransitionEdgeSensors2005}. Noise in different TESs is expected to be uncorrelated.

\begin{figure}
\includegraphics[width=1\columnwidth]{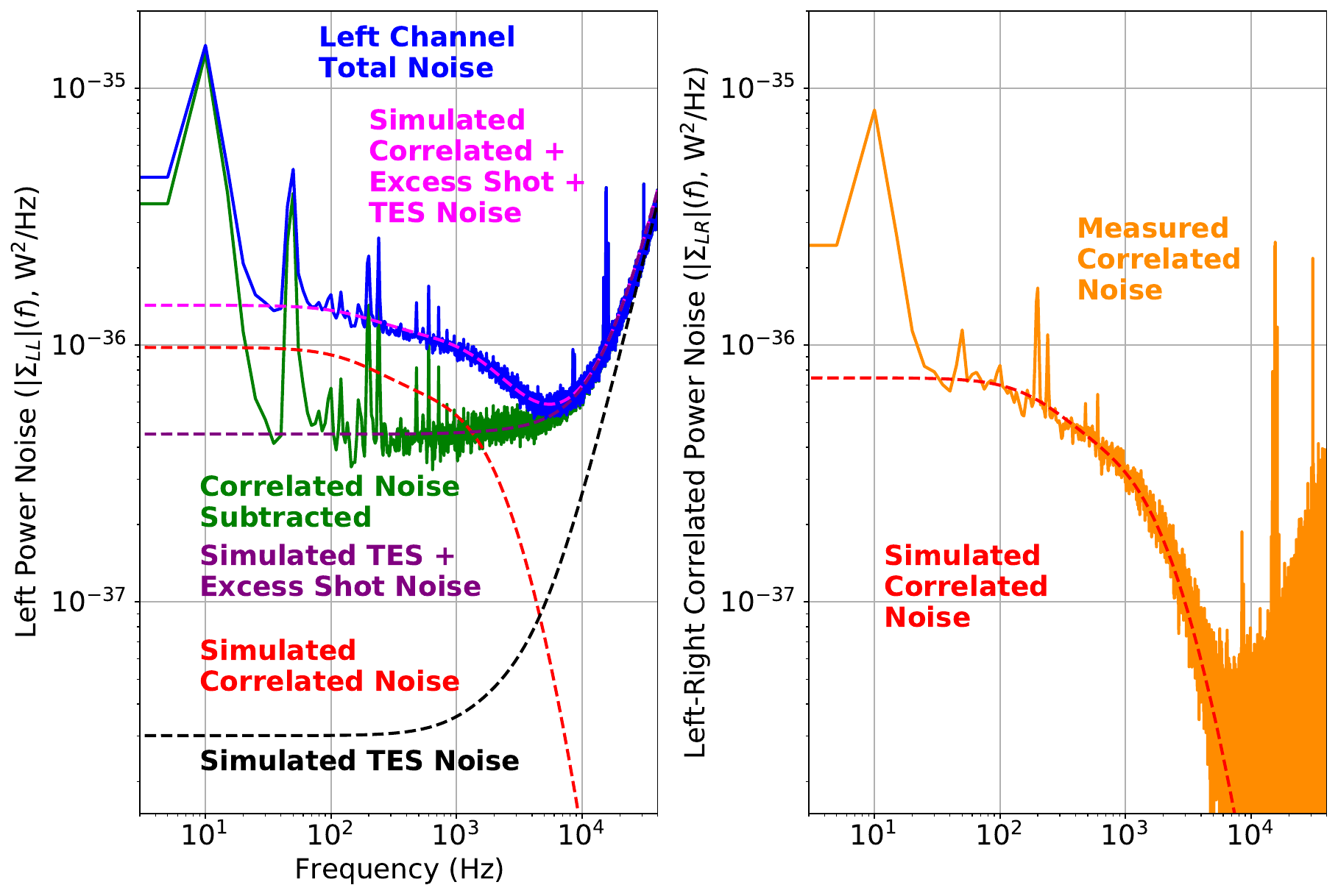}
\caption{\label{fig:noise} (Left) Left channel total noise (blue, $|\Sigma_{LL}|(f)$), ``uncorrelated'' noise (green) after subtracting modeled correlated noise (red, dashed, $a \Big|\frac{\partial P_{R}}{\partial E_p}\Big|^2(f)$), modeled TES noise (black, dashed, $M_L(f)$), excess shot noise + modeled TES noise (purple, dashed, $M_L(f) + U_L$),  and modeled noise including simulated excess correlated and shot terms (pink, dashed). (Right) Correlated noise (orange, $|\Sigma_{LR}|(f)$), fit to a phonon shot noise model (red, dashed, $a \Big|\frac{\partial P_{L}}{\partial E_p} \frac{\partial P_{R}}{\partial E_p}\Big|(f)$).}
\end{figure}

In our detector, noise in both channels is significantly above the modeled noise, and is correlated below several kHz (see Fig. \ref{fig:noise}). To elucidate the excess noise's source, we calculate the cross power spectral density (CSD) between the left and right channels for randomly-triggered time periods, cutting periods with high noise or above-threshold events. We convert this current CSD into the power domain by applying a responsivity model $\partial P/ \partial I(f)$ developed from the measured TES complex impedance $\partial V/ \partial I(f)$ \cite{ThreePoledPdI, watkinsThesis}. The on-diagonal elements of the power CSD $|\Sigma|(f)$ measure the total noise in each channel, while the off diagonal elements of the CSD estimate the correlated noise.

As the measured correlated noise rolls off very close to the measured athermal phonon collection pole, we model the CSD as the sum of three terms: modeled TES noise $M_{L,R}(f)$, uncorrelated shot noise $U_{L,R}$, and phonon shot noise $a \Big| \frac{\partial P_{L,R}}{\partial E_p}\frac{\partial P_{L,R}}{\partial E_p} \Big|(f)$, where $\frac{\partial P_{L,R}}{\partial E_p}(f)$ are the measured responses of the channels to phonon events during the photon calibration. 
%TC:ignore
\begin{widetext}
\begin{eqnarray}
    |\Sigma|(f) = \begin{bmatrix}
        |\Sigma_{LL}^2(f)| & |\Sigma_{RL}^2(f)| \\
        |\Sigma_{LR}^2(f)| & |\Sigma_{RR}^2(f)| 
    \end{bmatrix} = \begin{bmatrix}
        M_L(f) + U_L + a \Big|\frac{\partial P_{L}}{\partial E_p}\Big|^2(f) & a \Big|\frac{\partial P_{R}}{\partial E_p}\frac{\partial P_{L}}{\partial E_p} \Big|(f) \\
        a \Big|\frac{\partial P_{L}}{\partial E_p}\frac{\partial P_{R}}{\partial E_p}\Big|(f) & M_R(f) + U_R + a \Big|\frac{\partial P_{R}}{\partial E_p}\Big|^2(f)
    \end{bmatrix}
\end{eqnarray}
\end{widetext}
%TC:endignore

Figure \ref{fig:noise} compares this model to the measured noise, showing excellent agreement with only three degrees of freedom ($a, U_L, U_R$), and supporting the hypothesis that excess sub-threshold events create our observed excess noise. Sub-threshold LEE events that deposit energy instantaneously and locally within the sensor films would produce flat uncorrelated shot noise, while sub-threshold LEE events that generate a shared athermal phonon response would produce correlated shot noise with the observed frequency dependence. 

%TC:ignore
%\section{Discussion}
%TC:endignore

Our observations are consistent with excess events (i.e. LEE) above and below threshold creating backgrounds and shot noise respectively. With excess correlated noise and shared backgrounds, we observe phonon mediation and strong coupling to both channels. Similarly, we observe uncorrelated noise and singles that strongly couple to one channel and are inconsistent with substrate phonon events. If these excess noise terms are indeed LEE shot noise, LEE limits both resolution and backgrounds in our detector. Ultimately, mitigating this noise would allow us to achieve $\sim$ 60 meV sensitivities to phonon events, nearing the level needed to search for single optical phonons generated by dark matter interactions \cite{Knapen2018}.

Given the multiple classes of excesses, it seems plausible to attribute these events to multiple sources. For example, aluminum relaxation would naturally explain singles, while shot noise from GHz scale EMI bursts could dominate the uncorrelated noise. Likewise, different effects could dominate phonon backgrounds at different energy scales, e.g., high energy shared LEE might originate from the relaxation of radiation induced defects\cite{DefectLEE} or ``microfractures''\cite{astromFractureProcessesObserved2006} within the substrate, while excess correlated noise could be caused by the absorption of e.g. 40 meV photons emitted by the detector circuit board\cite{EssigBackgrounds}. We leave disentangling these hypothesized contributions to future work studying excess rates and noise over time and their scaling with properties of the detector and surrounding materials. 

In conclusion, we have demonstrated that two channel calorimeters provide key insights into excess noise and the LEE. Specifically, we show singles and uncorrelated noise are consistent with above and below threshold events in the aluminum sensor film. Additionally, these dual channel devices can be used to discriminate single LEE events that couple primarily to the sensor from events (and DM interactions) that couple to the detector phonon system, allowing for LEE to be partially discriminated in light DM searches. Understanding and disentangling these excesses will be key to unlocking meV-scale resolution phonon detectors and future highly-motivated searches for light dark matter. Our results may also be of interest to the superconducting quantum device community, who have long observed excess quasiparticles in their aluminum devices \cite{Mannila2021, Serniak2018, PlourdeQPs}.

%TC:ignore
\section{Supplementary Material}

See the supplementary material for details on our optimum filter based trigger scheme, our data quality cuts, the way in which we reconstruct the energy of triggered events, and the saturated events we see in our detector.

\section{Acknowledgments}

This work was supported in part by DOE Grants DE-SC0019319, DE-SC0022354, DE-SC0025523 and DOE Quantum Information Science Enabled Discovery (QuantISED) for High Energy Physics (KA2401032). This material is based upon work supported by the National Science Foundation Graduate Research Fellowship under Grant No. DGE 1106400. This material is based upon work supported by the Department of Energy National Nuclear Security Administration through the Nuclear Science and Security Consortium under Award Number(s) DE-NA0003180 and/or DE-NA0000979. Work at Lawrence Berkeley National Laboratory was supported by the U.S. DOE, Office of High Energy Physics, under Contract No. DEAC02-05CH11231. Work at Argonne is supported by the U.S. DOE, Office of High Energy Physics, under Contract No. DE-AC02-06CH11357. W.G. and Y.Q. acknowledge the support by the National High Magnetic Field Laboratory at Florida State University, which is supported by the National Science Foundation Cooperative Agreement No. DMR-2128556 and the state of Florida.

The authors have no conflicts to disclose. The data that support the findings of this study are available from the corresponding author upon reasonable request.
%TC:endignore

%\nocite{*}
\bibliography{aipsamp}% Produces the bibliography via BibTeX.

\appendix

\section{Triggering and Optimum Filtering}
\label{appendix:triggering_ofs}

For the analysis presented in this letter, we use optimum filters to trigger events from our recorded datastream and to measure the amplitude of triggered events.

As described in e.g. Ref.\cite{sunilThesis}, optimum (or matched) filters use pulse templates and noise spectra to construct a filter that optimally weights different frequencies to achieve the best performance possible for a given system (under certain conditions, e.g. Gaussian noise and linear response). It is important to note that optimum filters require the signal template be matched to the noisy signal being fit. For example, an optimum filter based on a shared template is not the best amplitude estimator for singles events (which have a different pulse shape). Therefore, different optimal filters need to be constructed for different purposes, e.g. estimating the amplitude of shared or single events.

Generically, we trigger by optimally filtering a recorded datastream, and recording the times when the optimally filtered trace exceeds a trigger threshold (defined in units of standard deviations above the mean, i.e. a 5$\sigma$ trigger) and reaches a local maximum. A triggered dataset consists of a list of such times, corresponding to events in the detector. In the analysis presented in this paper, we trigger events with four separate methods, as described in the text:
\begin{itemize}
    \item Calibration events (Figs. 2 and 4): we trigger an event when an optimally filtered logic signal associated with the laser firing exceeds a very large threshold (1000000$\sigma$), as the logic signal is essentially arbitrarily large compared to the noise. This trigger could be easily implemented with a simple threshold trigger, however, to maintain consistency across our analysis software stack, we use optimal filters for triggering calibration data.
    \item Shared events (Figs. 3 and 5 top): we sum together the recorded signal from the left and right detector channels, and construct an optimal filter for this summed trace by directly summing the shared pulse templates for the left and right channels (measured during the calibration) and by summing the noise in the left and right channels in quadrature. For these sums, we weight these two channels equally. We trigger events above a threshold of 5$\sigma$. Admittedly, this summing scheme is not truly optimal (we are currently implementing multi-channel ``N by M'' optimum filters, see e.g. Ref. \cite{kurinskyThesis}, which do perform this multi-channel triggering process optimally), but is sufficient for this analysis.
    \item Single events (Figs. 4 and 5 bottom): we trigger either left channel or right channel singles on a single channel optimum filter using a singles event template measured from our background dataset (i.e. the averaged events in the purple box in Fig. 3). We trigger events above 5$\sigma$.
    \item Random triggers (Fig. 6): to sample the noise, as well as to assess our cut efficiency and cross-check our energy resolution measurements, we additionally record triggers at randomly drawn times.
\end{itemize}

For the majority of our analysis, we require that triggers have a ``positive'' amplitude, i.e. that the current through our TESs decreases during the event, consistent with energy deposition in our TESs. However, random fluctuations in the noise may also exceed our trigger threshold, mimicking events created by true energy depositions within our sensors. As described in the text, we sample the rate of these ``noise triggers'' by looking for events with a ``negative'' (i.e. unphysical) amplitude, such that we only sample triggers from statistical fluctuations in our noise. This is equivalent to triggering on an inverted datastream, as in e.g. Ref. \cite{BULLKIDNegTrigs}. 

After triggering, we use optimal filters to estimate the amplitude of triggered events, again tailoring the signal template used to the signal being measured (i.e. estimating shared amplitudes using a shared template optimal filter). In our analysis pipeline, we construct these optimum filters such that the output is the height of the pulse in units of current through our TES, a quantity proportional but not equal to the energy of the event in the linear regime of the TES. In addition to recording the best fit amplitude, we also record the $\chi^2$ goodness of fit parameter, as well as a ``low frequency" goodness of fit statistic $\chi^2_{LF}$, which calculates the goodness of fit for frequencies below a cutoff frequency (in our case 50 kHz), above which there is minimal template information, lowering the sensitivity of this statistic to pulse shape or noise differences.

\section{Data Quality Cuts}
\label{appendix:cuts}

We apply standard quality cuts to our dataset to ensure the detector is operating stably and that abnormal events minimally impacts our data quality (e.g. periods of high vibration induced noise, event pileup, altered TES response shortly after the detector is saturated). 

\begin{figure}
\includegraphics[width=1\columnwidth]{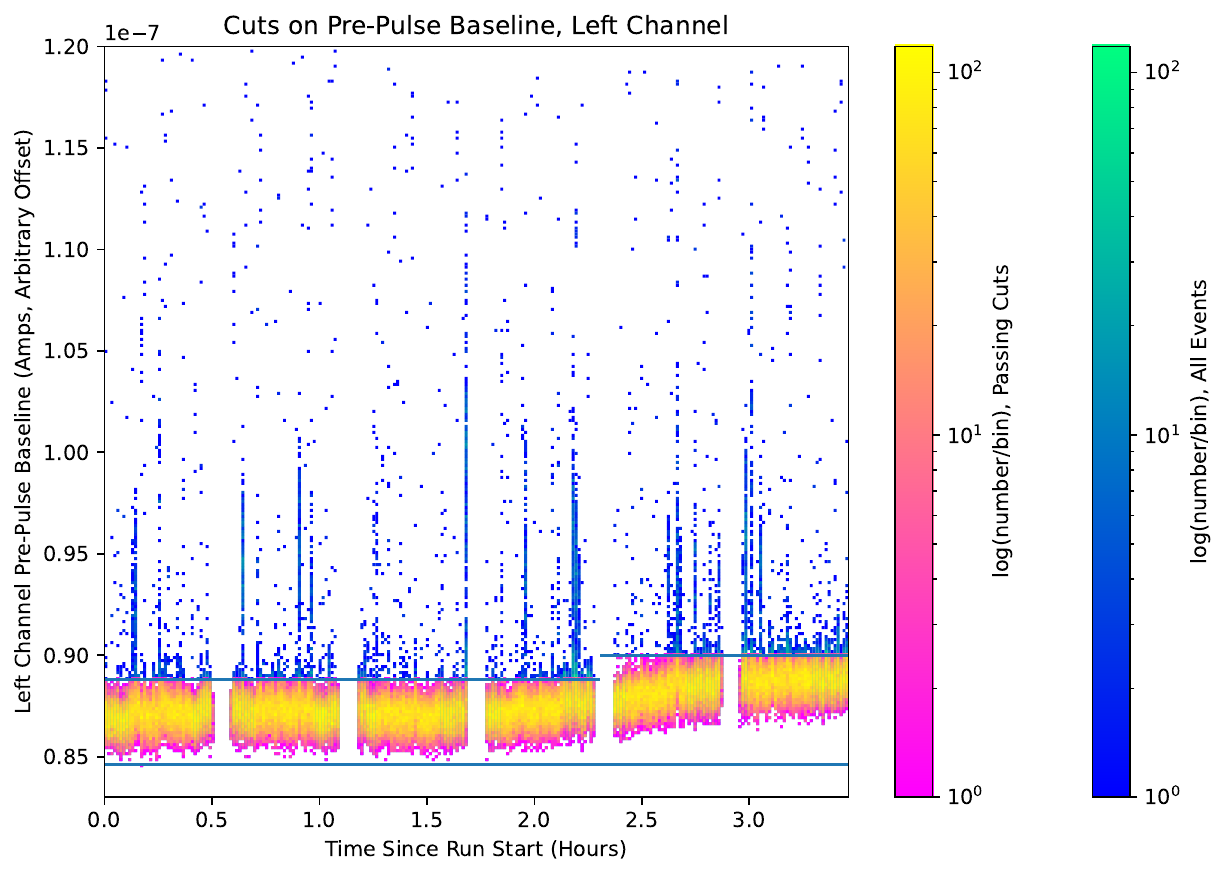}
\caption{\label{fig:cut_baseline} Plot showing the left channel pre-pulse baseline cut (see text). Blue lines show the levels of the cuts, pink/yellow points show passing events, and blue/green points show failing events. Three hours of data is taken in 6 30 minute periods, with detector characterization ($\partial I/ \partial V$, IV measurements) taken in between (blank periods). Vertical lines of events are triggers during periods when the device temperature is elevated for $\sim$ 10 seconds following a high energy events. The shift in the baseline over time is from an unknown source. Plot is cropped for clarity, points for events following high energy events continue up to $\sim2.1\times10^{-7}$ amps.}
\end{figure}

\begin{figure}
\includegraphics[width=1\columnwidth]{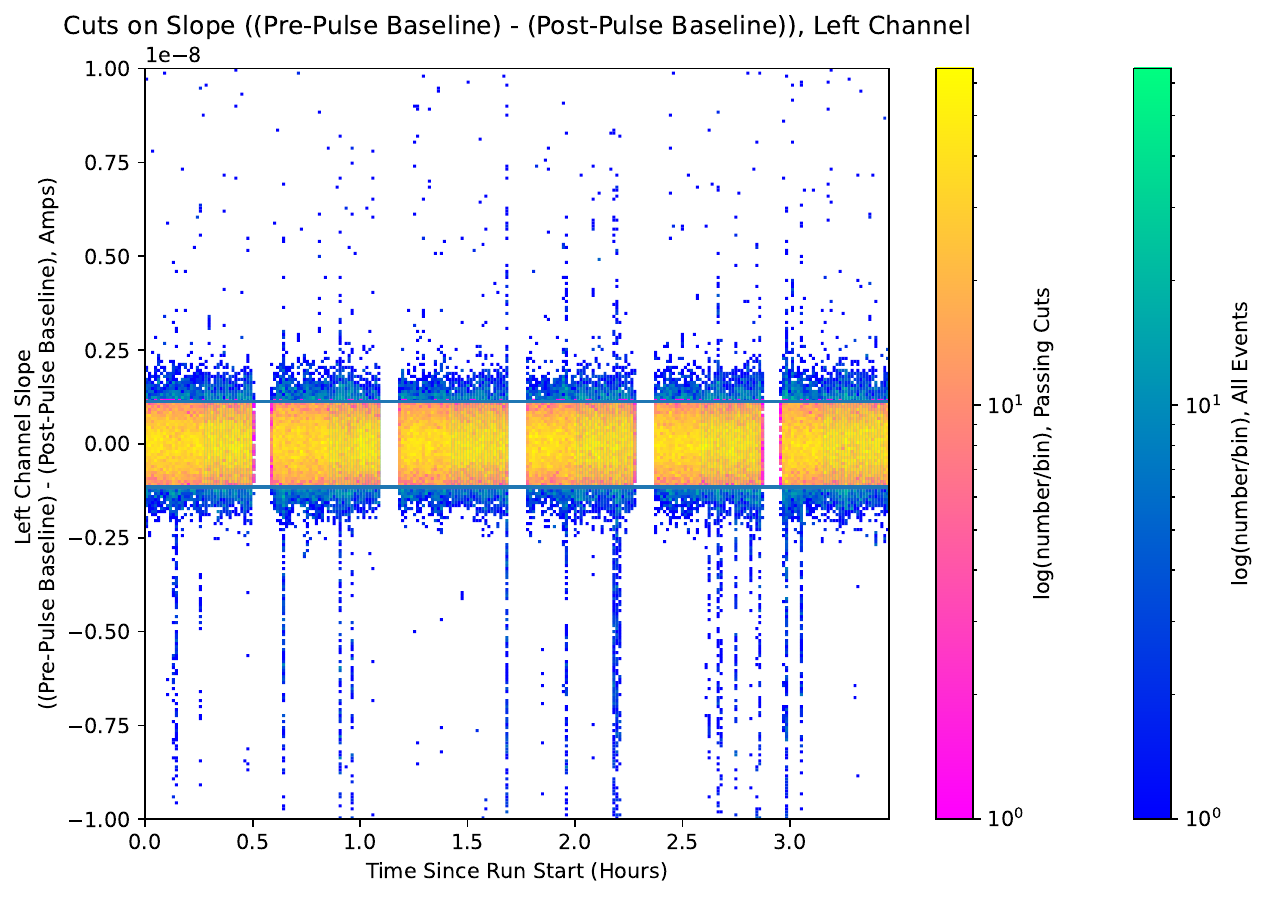}
\caption{\label{fig:cut_slope} Plot showing the left channel ``slope'' (pre-pulse baseline minus post-pulse baseline) cut (see text). Blue lines show the levels of the cuts, pink/yellow points show passing events, and blue/green points show failing events. Three hours of data is taken in 6 30 minute periods, with detector characterization ($\partial I/ \partial V$, IV measurements) taken in between (blank periods). Vertical lines of events are triggers during periods when the device temperature is elevated for $\sim$ 10 seconds following a high energy events. Note that saturated events fail this cut by default, they are passed manually later in the analysis pipeline.}
\end{figure}

\begin{figure}
\includegraphics[width=1\columnwidth]{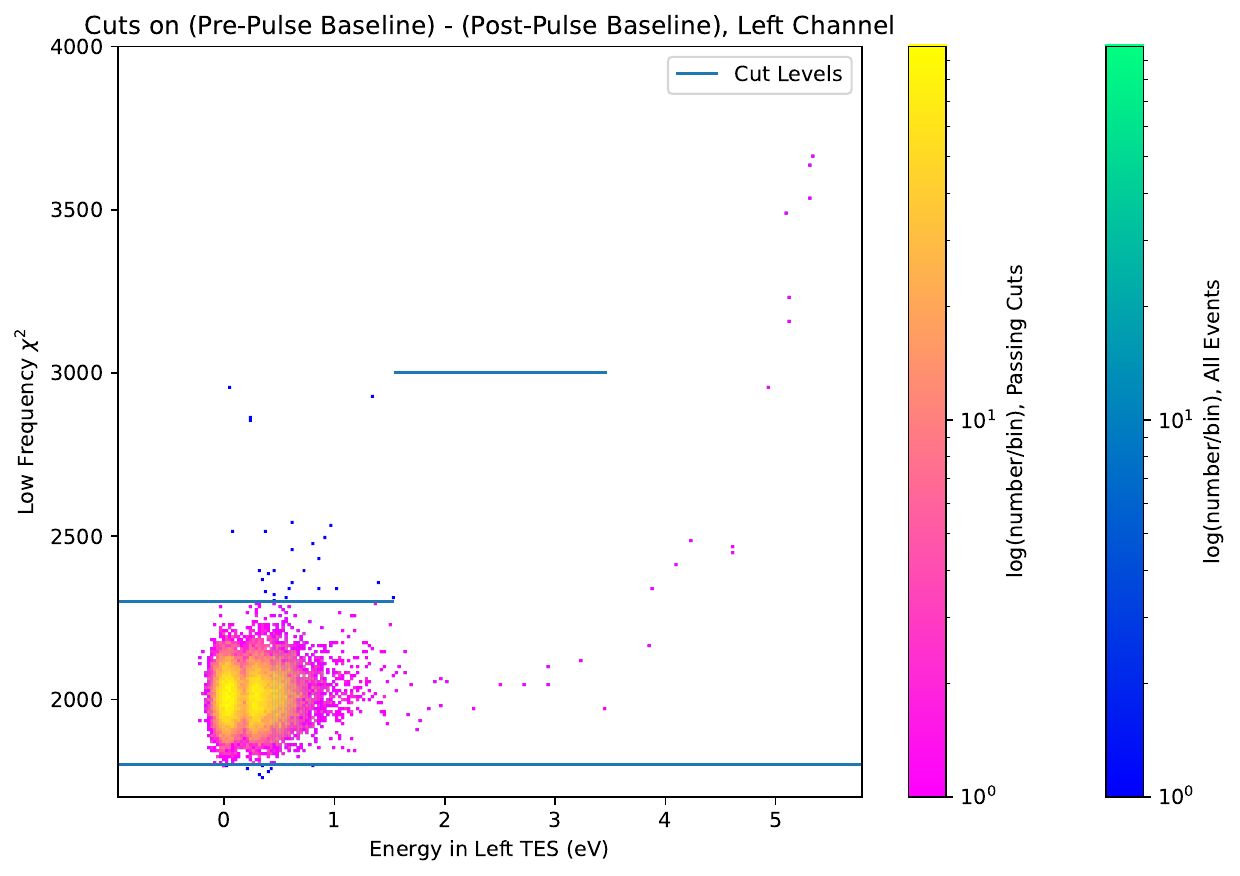}
\caption{\label{fig:cut_chi2} Plot showing the left channel $\chi^2_{LF}$ cut (see text). Blue lines show the levels of the cuts, pink/yellow points show passing events, and blue/green points show failing events. Only events which trigger on the channel sum trigger and which pass the baseline and slope cuts are shown for clarity. Shared events (see text) curve up towards higher $\chi^2_{LF}$ values as they become more and more saturated. As the channel sum trigger triggers on right channel singles as well as shared events and left channel singles, these right channel singles appear as a quasi-Gaussian group of events centered near zero energy in the left channel. Cut events somewhat above the main band of events at low energies are pileup events. The energy reconstruction for the x axis of this figure uses a shared template.}
\end{figure}

To all data presented in this paper, we apply three ``standard'' quality cuts:
\begin{itemize}
    \item Pre-pulse baseline cut (see Fig. \ref{fig:cut_baseline}): we average the current running through the TES between 10 and 2 ms before an event is triggered, and only accept events for which this average falls beneath a time dependent threshold. As this cut only considers the detector performance before an event, it is a fully pulse shape independent cut. We set this threshold independently for each 30 minute datataking period (between which we characterize the detector with IV sweeps and $
    \partial I/\partial V$ measurements) such that the passage fraction is roughly equal (in this case at around 95\%). This cut primarily rejects events which occur shortly ($\sim$ 10s) after a high energy event occurs in the detector, temporarily elevating the device temperature and raising the pre-pulse baseline. Additionally, it partially rejects pileup. We notice long ($\sim$ hour) period shifts in the average pre-pulse baseline from an unknown source. We do not attempt to reject these shifts, as they change the device bias power by roughly 1 $\%$, minimally shifting the TES response. In other devices, we have noticed shorter ($\sim$ 5 min long) periods where the the detector baseline is elevated that we attribute to microphonic (i.e. vibration induced) heating, however, this detector appears to be less sensitive to this heating source for unclear reasons. 96.4 \% of randomly triggered events pass this pre-pulse baseline cut.
    \item ``Slope'' cut (see Fig. \ref{fig:cut_slope}): we subtract a ``post-pulse'' baseline (average current through the device between 2 and 10 ms after the event triggers, after the detector has returned to essentially the baseline level) from the pre-pulse baseline to to calculate the ``slope'' of the trace. We then cut events for which this slope quantity falls below or above a threshold. This cut is at least partially redundant with the pre-pulse baseline cut, and largely serves to cut post-pulse event pileup (i.e. multiple events in one trace) when applied with the pre-pulse baseline cut. Note that highly saturated events fail this cut (as they are still at an elevated baseline 2 ms after the initial trigger). We therefore pass all high amplitude (saturated) events. 81.5 \% of all randomly triggered events pass this slope cut.
    \item $\chi^2_{LF}$ cut (see Fig. \ref{fig:cut_chi2}): our optimum filter automatically calculates a goodness of fit parameter $\chi^2_{LF}$ for each event, which compares the difference between the (Fourier transformed) fit shared template and the measured signal in each frequency bin below a cutoff frequency (in our case 50 kHz, above which there is minimal signal information). At higher amplitudes, our pulse shape deviates from the nominally expected low amplitude pulse shape as the TES becomes more and more non-linear, until it changes completely when the device saturates. We therefore set an amplitude dependent cut which cuts events above a threshold which is lower at low amplitude than at higher amplitudes.

    This cut serves several purposes. It removes events with an unexpected shape, most frequently pileup events or events which trigger at incorrect times (i.e. ``trigger echos''). The $\chi^2_{LF}$ cut is therefore effectively a pulse shape cut, but does not cut on specific quantities, e.g. rise times or height to integral ratio. To ensure that this cut does not remove otherwise good events with an unexpected pulse shape (e.g. singles) we have tested the effect of removing this cut on the analysis. After removing this cut, there was no significant change to the distribution of shared or single events, or to their average pulse shape. We apply the $\chi^2_{LF}$ cut in our final analysis to primarily remove pileup and incorrect time triggers.

    99.6 \% of all randomly triggered events pass the $\chi^2_{LF}$ cut.
\end{itemize}
These cuts are similar to those applied in previous analyses by our group, e.g. \cite{AnthonyPetersen2024, CPDLimits}. For all parts of this analysis, we require events to pass all three of these cuts on both channels. All three cuts are ``efficiency'' rather than ``time acceptance'' cuts.

To minimally bias our dataset, we design our cuts for high passage fraction. To estimate the passage fraction of low energy events (which are of most interest for our LEE analysis) we calculate the passage fraction of randomly triggered events, which are effectively very low (zero) energy events. We pass 79.5$\%$ of randomly triggered events and 75.9$\%$ of triggered events. We expect the triggered event passage fraction to be slightly smaller than for randoms, due to a small fraction of events which trigger at the incorrect time.

Broadly, the conclusions of are study are unaffected by applying or removing these cuts. For example, without these cuts, we see pileup events or incorrect trigger time events polluting our pulse shape measurements, and we see slight changes to measured pulse shapes and noise spectra as our TESs shift their response as they move higher in their transitions.

In addition to these ``standard'' quality cuts, we apply two different cuts in different phases of the analysis. As described in the text, we compare the $\chi^2$ statistic for shared and single fits to discriminate between shared and single events for Figure 5, so that the top plot shows selected shared events and the bottom shows selected single events. For the noise analysis (Fig. 6), we select traces for which the largest optimally filtered amplitude in the trace is negative (as in e.g. Ref.\cite{BULLKIDNegTrigs}) to select events which do not have a sufficiently large true energy deposition event to bias the measured noise. To ensure that our ``no event'' cut was working as expected, we studied the noise we measured as a function of the strictness of this cut, and found that as long as the very highest energy ($> 20 \sigma$) events were removed, the resulting noise spectrum was statistically identical regardless of the strictness of our cut. This is consistent with our noise being dominated by a large number of sub-threshold events, as proposed in the main text.  

\section{Energy Reconstruction}
\label{appendix:energy_reconstruction}

Our TES-based phonon sensors are not perfectly efficient at converting energy deposited into the detector phonon system into energy in the TES. Some phonon energy is lost along the way, through e.g. absorption of phonons into non-instrumented films on the detector, through the physics of quasiparticle creation, or quasiparticle trapping within the aluminum phonon collection fins. As at the energies relevant to our detectors this loss is essentially energy dependent, we can relate
\begin{eqnarray}
    E_{\mathrm{TES}} = \epsilon_\mathrm{PCE} E_\mathrm{phonon}
\end{eqnarray}
where $E_{\mathrm{TES}}$ is the energy absorbed in the TES, $E_\mathrm{phonon}$ is the energy deposited in the phonon system, and $\epsilon_\mathrm{PCE}$ is the ``phonon collection efficiency'' (PCE). As this letter discusses events which occur both in the phonon system and which couple to TESs without phonon mediation, we primarily show results in units of $E_{\mathrm{TES}}$ (``Energy absorbed in Left/Right Channel''). We can measure this energy because the response of the current flowing through a TES to a given power input is a well studied and modeled. By integrating the reconstructed power put into the detector, we can convert this power signal into an energy which was absorbed in a given TES in a given event.

\subsection{Energy Reconstruction Method}

In our analysis, we compensate for our room temperature amplification, SQUID feedback resistance and turns ratio such that we can reconstruct the current flowing through our TES from the voltage signal recorded in our DAQ at room temperature. Using these current domain signals, we then use optimum filters to fit the amplitude of these pulses, in units of current. From these current domain event amplitudes, we convert into event energies as in Ref. \cite{TESVeto} by calculating a factor $\frac{\partial E}{\partial I}$.
\begin{eqnarray}
    \frac{\partial E}{\partial I} = \int T_p(t) dt =  \int {\mathcal{F}}^{-1} \bigg( \frac{\partial P}{\partial I}(f) \mathcal{F}(T_i(t)) \bigg) dt
\end{eqnarray}
Here, $T_i(t)$ is the time and current domain template used in the optimum filter, $T_p(t)$ is this same template converted into the power domain, and $\frac{\partial P}{\partial I}(f)$ is the responsivity of the TES\cite{irwinTransitionEdgeSensors2005, ThreePoledPdI, watkinsThesis}.

We model the $\frac{\partial P}{\partial I}(f)$s of our TESs following the approach in Ref. \cite{ThreePoledPdI}, assuming the TES has exactly three dynamical poles. To construct this model, we measure the TES's dynamic and static properties through complex impedance ($\frac{\partial V}{\partial I}(f)$, in which we modulate the TES bias with a small square wave) and IV measurements respectively.

To cross check this calculation of $\frac{\partial E}{\partial I}$, we have performed a simpler ``zero-frequency'' calculation (as in Ref. \cite{AnthonyPetersen2024}), which simply integrates
\begin{eqnarray}
    \frac{\partial E}{\partial I} \approx \frac{\partial P}{\partial I} \int dt T_I(t) dt \\
    = (2 I_{\mathrm{TES}} R_\mathrm{load} - V_\mathrm{bias}) \int dt T_I(t) dt
\end{eqnarray}
which yields consistent results from the calculation above within about 10$\%$.

\subsection{Phonon Collection Efficiencies and Notes on TES Performance Degradation Through Tungsten Etching}

Under this paradigm, our calibration allows us to measure $\epsilon_\mathrm{PCE}$ for each channel, as we know the true energy of the phonon event (e.g. 3.061 eV for calibration events in the first peak of the spectrum in Fig. \ref{fig:calibration} insert) and as we can calculate the energy absorbed in each TES channel during these events.

\begin{figure}
\includegraphics[width=1\columnwidth]{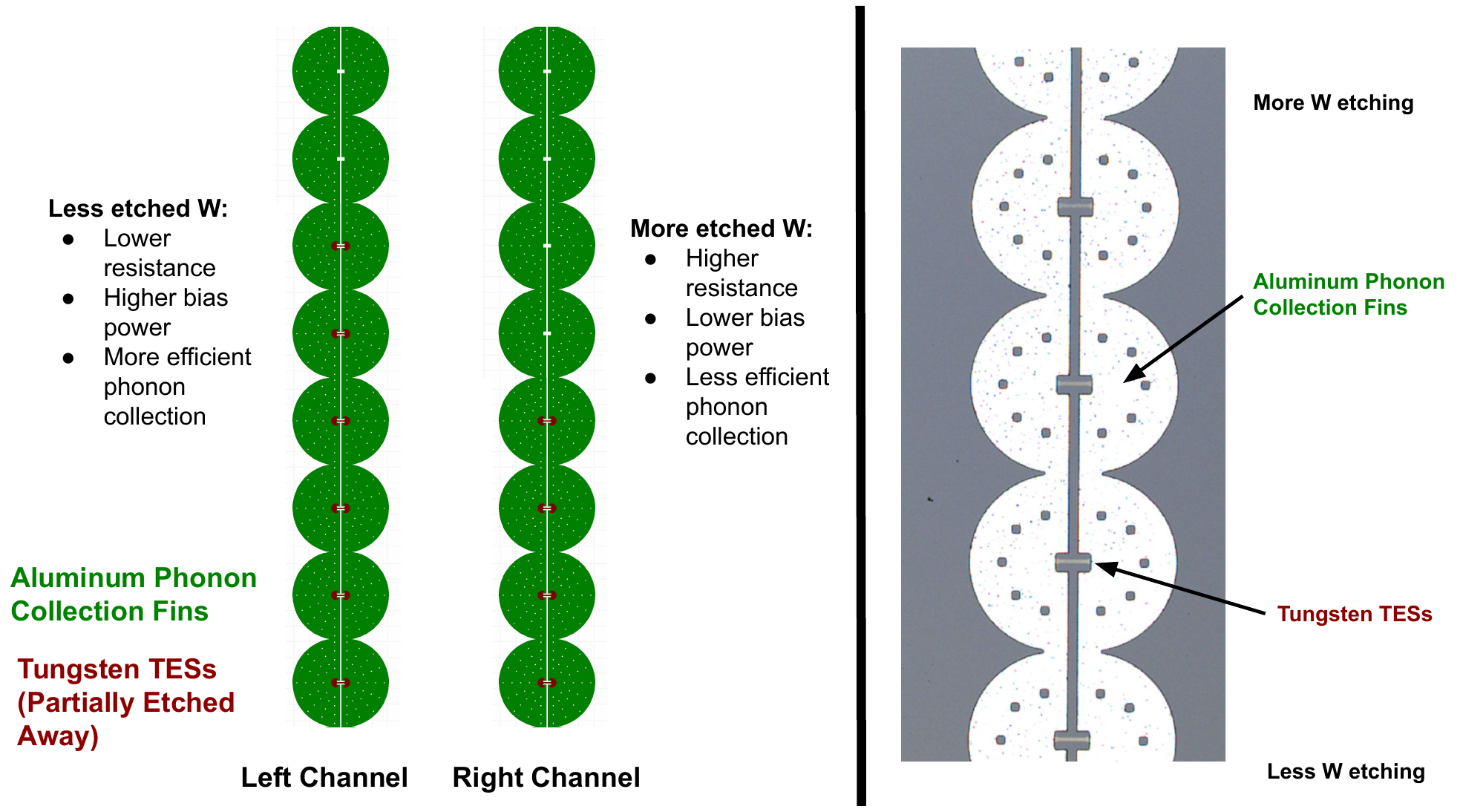}
\caption{\label{fig:etched_w} (Left) Sketch of the model we use to explain the difference in performance between the two channels. The left channel has had the tungsten etched away to a lesser degree, leading to a lower normal resistance, a higher bias power in transition, and more phonon collection. (Right) Microscope image of a device of a similar design from the same wafer. Note the lighter (more etched) grey-yellow tungsten TESs at the top of the column of devices compared to the TESs at the bottom. In other locations on the wafer, we observed both completely etched away tungsten and essentially intact tungsten.}
\end{figure}

We measure $\epsilon_\mathrm{Left} = 0.164 \pm 0.001$ and $\epsilon_\mathrm{Right} = 0.125 \pm 0.001$ for the left and right channels respectively. This measurement is consistent with our measured device normal resistances (Left: 1.075 $\Omega$, Right: 1.148 $\Omega$) and bias powers (Left: 4.180 fW, Right: 1.726 fW) if we assume that fewer of the TESs are etched away in the left channel compared to the right channel. A less etched channel would have a lower normal resistance, a higher bias power, and collect phonons more efficiently. See Fig. \ref{fig:etched_w} for a sketch of this model.

As the phonon collection efficiencies are measured during the photon calibration, the difference in relative $\epsilon_\mathrm{PCE}$ is visible in Fig. 2. For example, photons in the 2 photon peak deposit around 1 eV of energy in the Left TES ($2E_\gamma\epsilon_\mathrm{Left} \approx 2 \times 3.061$ eV $\times 0.164 \approx 1$ eV), while the same 2 photon events deposit around 0.8 eV in the Right TES ($2E_\gamma\epsilon_\mathrm{Right} \approx 2 \times 3.061$ eV $\times 0.125 \approx 0.77$ eV). The same effect is also visible in the background dataset (Fig. 3). Events which deposit 3 eV in the right channel deposit about 4 eV in the left channel, in line with their relative phonon collection efficiencies ( (3 eV)/(4 eV) $\approx \epsilon_\mathrm{Right}/\epsilon_\mathrm{Left}$).

\subsection{Phonon Energy Reconstruction: Inverse Variance Weighting}

To most accurately estimate the energy a given event deposits in our phonon system, we use inverse variance weighting to combine the signals from the left and right channels. This method is the same as we use in Ref. \cite{TESVeto}.

For a given energy absorbed in each channel for a phonon event $E_\mathrm{TES}$, we could estimate the energy deposited in the phonon system $E_P$ using the phonon collection efficiency $\epsilon$ as
\begin{eqnarray}
    E_{P,L} = E_{TES,L} \epsilon_L \\
    E_{P,R} = E_{TES,R} \epsilon_R
\end{eqnarray}
We combine these two signals by weighting by the inverse of the variance in each channel, i.e. $1/\sigma^2_{P,L}$ or $1/\sigma^2_{P,R}$, where $\sigma_{P,L}$ and $\sigma_{P,R}$ are the uncertainties in the $E_{P,L}$ and $E_{P,R}$ respectively. This yields our estimated phonon energy $E_P$ considering the response in both the left and right channels.
\begin{eqnarray}
    E_P = \frac{1}{\frac{1}{\sigma^2_{P,L}} + \frac{1}{\sigma^2_{P,R}}} \bigg(\frac{E_{P,L}}{\sigma^2_{P,L}} + \frac{E_{P,R}}{\sigma^2_{P,R}} \bigg)\\
    = \frac{1}{\frac{1}{\sigma^2_{P,L}} + \frac{1}{\sigma^2_{P,R}}} \bigg(\frac{E_{TES,L} \epsilon_L}{\sigma^2_{P,L}} + \frac{E_{TES,R} \epsilon_R}{\sigma^2_{P,R}} \bigg)
\end{eqnarray}

We plot this quantity in the inset in figure 2, and use it to measure our phonon energy resolution by fitting a Gaussian to the zeroth photon peak.

\subsection{Template Dependence}

\begin{figure}
\includegraphics[width=1\columnwidth]{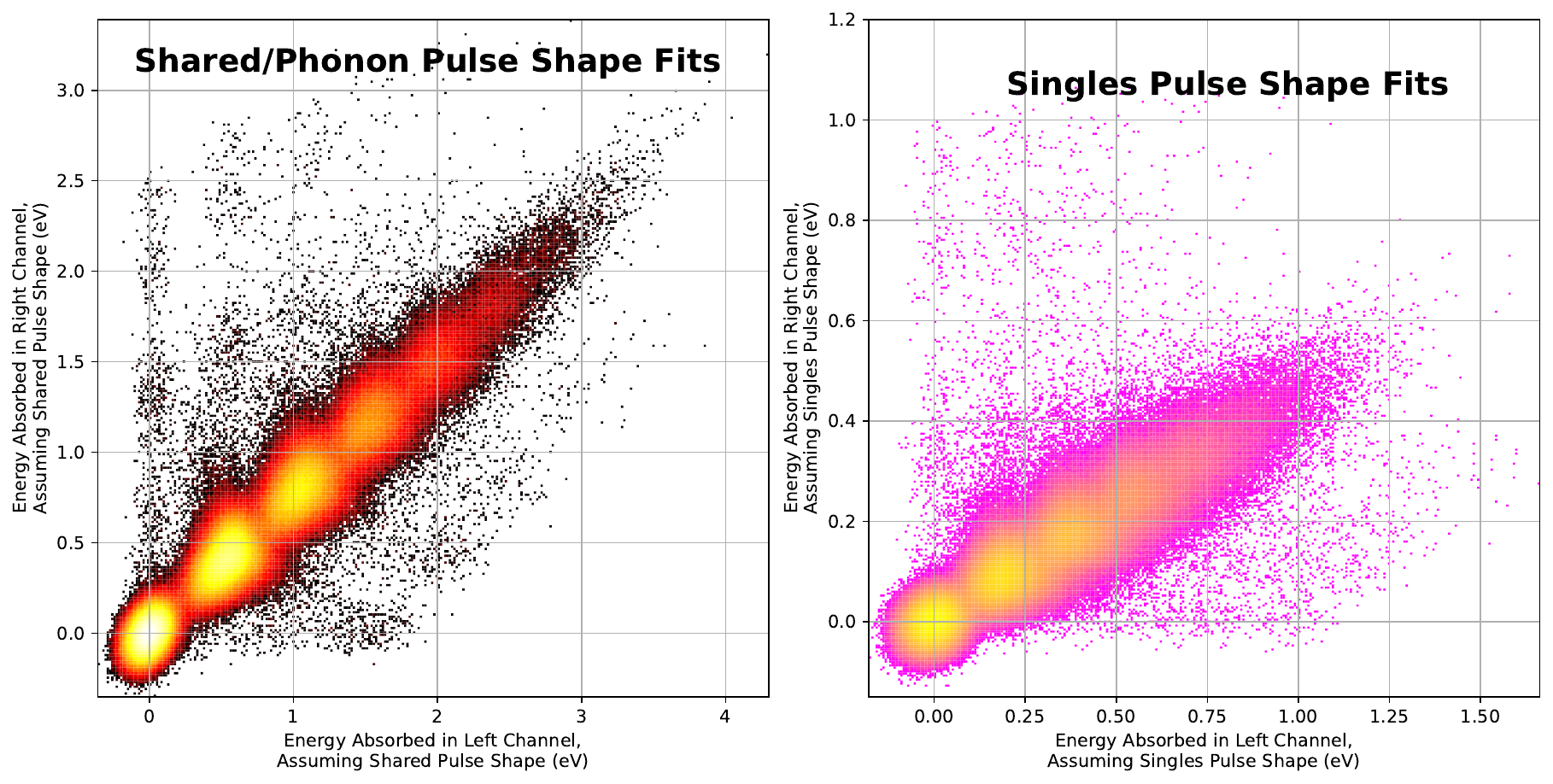}
\caption{\label{fig:of_comparison} A comparison of reconstructed energies using shared and singles templates to fit the calibration dataset.  (Left) Fits using a shared/phonon template (i.e. upper right figures in Figs. 2 and 3). This plot is the same as the main panel in Fig. 
2 (Right) The same data, fit using optimum filters based on singles templates. Note the difference in energy scale for both the shared and singles events between the two subfigures. Also note that in the right panel, the two direct hit/singles bands extend to roughly the same energy, consistent with the same underlying physical mechanism (photon absorbtion in the aluminum phonon collection fins of the QETs).}
\end{figure}

Finally, it is important to emphasize that an optimum filter which uses a pulse shape template that is incorrect will not accurately estimate the energy of these events. For example, an optimum filter which uses a shared event template will not accurately estimate the energy of singles events (and visa versa). This leads to the singles or direct hit events appearing at inaccurate energies in Figs. 3 and 2, as the analysis for these figures assumes a shared pulse shape. See Fig. \ref{fig:of_comparison} for an example of this effect, analyzing the calibration dataset with both shared and singles template based optimuim filters.

For clarity, in the following figures we use (as noted in their captions):
\begin{itemize}
    \item Figures 2 and 3: Shared/phonon templates
    \item Figure 5 top (and top inset): Shared/phonon template
    \item Figure 5 bottom (and bottom inset): singles template
\end{itemize}

\section{Saturated Events}
\label{appendix:saturated}

For small events, TESs operate as essentially linear devices thanks to strong feedback and a steep resistance to temperature relationship around their operating point. However, sufficiently large events may deposit enough energy in the TES to drive them out of the linear region of their transition, into the normal regime. Once normal, the measured current flowing through the TES is essentially independent of the TES temperature, giving these events a flat top. In this regime, it is very difficult to accurately reconstruct the energy deposited in the TES, as it is almost entirely dissipated through highly nonlinear electron-phonon coupling and as the current through the TES no longer strongly related to the TES temperature.

These events are often called ``saturated events,'' given trying to estimate the energy deposited for these events using standard linear techniques does not exceed a certain level. The energy scale at which these events start to saturate our TES array is very roughly the bias power of the channel (left: $\sim$ 4 fW, right: $\sim$ 2 fW) multiplied by the time scale of the event (see Fig. 2 top right, very roughly left: 200 $\mu$s, right: 300 $\mu$s). In our detector, we observe that events that deposit more than 5 eV in the left channel or 3.5 eV in the right channel enter this saturation regime (assuming a phonon template). This occurs for events which deposit more than about 30 eV in the detector phonon system. 

Many of these events likely deposit far more than 30 eV in the detector phonon system. For example, cosmic rays and gamma rays from background radioactivity are expected to deposit hundreds of keV or even MeV in our silicon substrate \cite{MartinisSavingQubits}. We assume that muons and other cosmic rays reach our detector at the widely cited flux of about 1 Hz/cm$^2$, and observe 0.051 Hz of saturated events in our detector. We attribute the discrepancy between this and the nominally expected 0.0167 Hz rate from cosmic rays to several factors:
\begin{itemize}
    \item Cosmic ray interactions with materials around the detector (for example the PCB we use to read out the detector, or the copper detector housing) may cause secondary showers or fluorescence\cite{EssigBackgrounds} which deposits much more than 30 eV of energy in our detector. 
    \item Radiogenic backgrounds (from e.g. materials in the cryostat, concrete lab walls, etc.) will also contribute to the saturated event rate. If we assume that we have 1000 DRU (daily rate units, 1 DRU = 1 event per keV per kg of detector per day) of backgrounds, and that our radiogenic backgrounds extend up to 1 MeV in energy, we would expect on order 0.0025 Hz of backgrounds. Similarly to cosmic rays, interactions with materials around the detector may increase this rate.
    \item We expect small contributions from the phonon-coupled LEE spectrum, which continues above 30 eV \cite{adariEXCESSWorkshopDescriptions2022}.
\end{itemize}

This saturated rate is also consistent with the rate we measured with two different detectors in the same fridge in Ref. \cite{AnthonyPetersen2024}.

\section{Comparison with Anthony-Petersen et. al. 2024}

Here, we compare the shared LEE rate with the rate observed in Ref. \cite{AnthonyPetersen2024}. As Ref. \cite{AnthonyPetersen2024} did not directly calibrate their devices, we assume an effective collection efficiency of $\epsilon = 0.25$ to convert between ``energy absorbed in the TES channel'' (as originally published) and ``energy deposited in the detector phonon system'' (as we show here, to compare to our results).

\begin{figure}
\includegraphics[width=1\columnwidth]{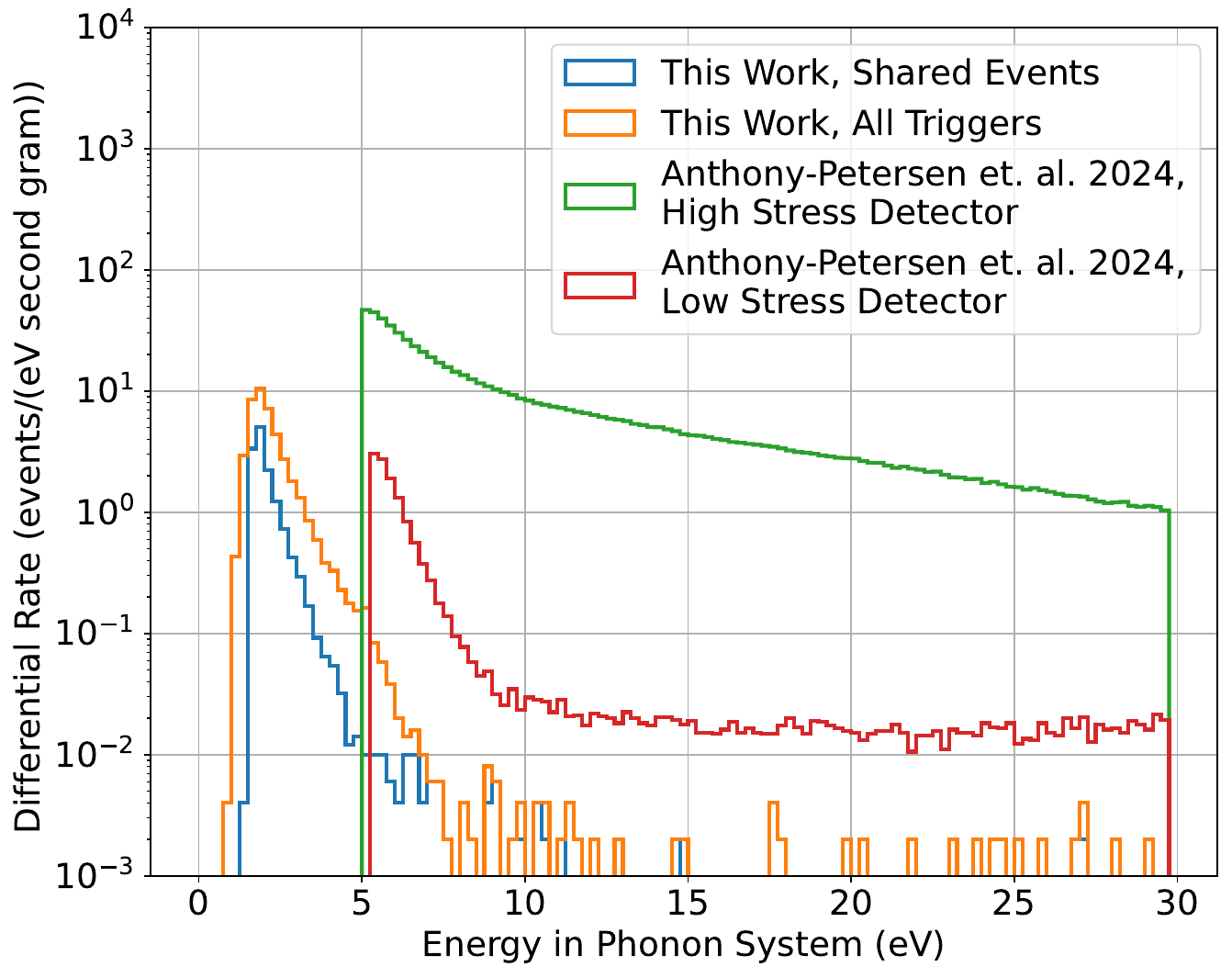}
\caption{\label{fig:r17_comparison} A comparison between the observed backgrounds in Ref. \cite{AnthonyPetersen2024} and this work. ``All triggers'' is all events passing the core analysis cuts (baseline, slope, low frequency $\chi^2$), including single events, while ``Shared Events'' includes the single-shared $\delta \chi^2$ cut. We assume a phonon collection efficiency of 0.25 for both detectors in Ref. \cite{AnthonyPetersen2024} to reconstruct a phonon spectrum, and use the last dataset taken in the data taking campaign. The ``High Stress'' detector was glued to a copper mount using GE Varnish, and the ``Low Stress'' detector was suspended from wire bonds, as the detector in this work was.}
\end{figure}

We see that the rate of background events is lower in the detector used in this work, for both all (single and shared) events and shared events only, and compared to both the ``high stress'' (glued) and ``low stress'' (hanging) detectors. These two datasets were taken at comparable times after cooldown, suggesting that variations in background rate with time would not easily explain the observed discrepancy. We instead posit that for unknown reasons, the LEE varied in rate between the sets of devices used for these two studies.

%\bibliography{aipsamp}% Produces the bibliography via BibTeX.

\end{document}